\documentstyle[12pt]{article}

\begin{document}

\begin{titlepage}
\begin{flushright}
RAL-TR-97-005\\
hep-th/9701120
\end{flushright}

\begin{center}
{\Large\bf Physical Consequences of Nonabelian Duality in the Standard Model}\\
\vspace{1cm}
{\large CHAN Hong-Mo\footnote{chanhm\,@\,v2.rl.ac.uk}}\\
{\it Rutherford Appleton Laboratory,\\ Chilton, Didcot, Oxon 0X11 0QX,
   U.K.}\\
\vspace{.5cm}
{\large TSOU Sheung Tsun\footnote{tsou\,@\,maths.ox.ac.uk}}\\
{\it Mathematical Institute, Oxford University, \\24-29 St.\ Giles', Oxford
   OX1 3LB, U.K.}\\
\end{center}
\begin{abstract}
Possible physical consequences of a recently discovered nonabelian
dual symmetry are explored in the standard model.  It is found that 
both Higgs fields and fermion generations can be assigned a natural 
place in the dual framework, with Higgs fields appearing as frames 
(or ``$N$-beins'') in internal symmetry space, and generations
appearing as spontaneously broken dual colour.  Fermions then occur 
in exactly 3 generations and have a factorizable mass matrix which 
gives automatically one generation much heavier than the other two.  
The CKM matrix is the identity at zeroth order, 
but acquires mixing through higher loop corrections.  Preliminary 
considerations are given to calculating the CKM matrix and lower 
generation masses.  New vector and Higgs bosons are predicted.
\end{abstract}

\end{titlepage}

\clearpage

\baselineskip 16pt

\section{Introduction}

The long-standing interest in electric-magnetic duality 
\cite{Dirac}-\cite{Deser} and its nonabelian generalizations \cite{Wuyang},
\cite{Lubkin}-\cite{Chanstsou} has seen an active revival in the last 
few years \cite{Seiten}-\cite{Chanftsou1} and much effort has been 
devoted in finding their physical consequences.

In a previous paper \cite{Chanftsou} it was shown that
Yang--Mills theories are symmetric under a generalized dual
transform which reduces to the familiar Hodge star operation
in the abelian case.  The purpose of the present paper  is to
examine what physical consequences this dual symmetry might have
when applied to the standard model with gauge symmetry
$su(3) \times su(2) \times u(1)$, which seems to embody all
presently known facts in particle physics apart from gravity.

We note that in contrast to many other approaches to duality
adopted in the literature which aim at extending the standard
model to a larger theory making use of supersymmetry and higher
dimensions of space-time and/or constituents of matter (such as
strings and membranes), we choose here to aim for economy.  In
other words, instead of looking `beyond the standard model' as
is often done, we shall remain strictly within the standard
model framework in 4 space-time dimensions, and ask merely
whether, within this framework, the recently discovered
generalized nonabelian dual symmetry can lead to physical 
consequences which are as yet unknown or unexplored.

Now although this generalized dual symmetry has strictly
speaking been established only for classical fields, we wish
to show that when supplemented by some general known facts in
quantum field theories, plus some seemingly reasonable assumptions
special to our treatment, the symmetry when applied to the standard
model can lead to predictions of quite considerable interest.  
Before studying these for the standard model in detail, however, let 
us first examine duality for Yang--Mills theory in general terms for 
indications in which areas physical consequences may be expected
to arise.

We recall the generalized dual transform proposed in our earlier paper 
\cite{Chanftsou}:
\begin{eqnarray}
\lefteqn{
\omega^{-1} (\eta(t)) \tilde{E}_\mu [\eta|t] \omega(\eta(t)) } \nonumber \\
&& =-\frac{2}{\bar{N}} \epsilon_{\mu\nu\rho\sigma} \dot{\eta}^\nu \int
\delta \xi ds E^\rho [\xi|s] \dot{\xi}^\sigma (s) \dot{\xi}^{-2} (s)
\delta (\xi(s)-\eta(t)),  \label{dualtransf}
\end{eqnarray}
which was expressed in terms of some loop space variables $E_\mu[\xi|s]$
and its dual $\tilde{E}_\mu[\xi|s]$ describing the gauge field.  The 
actual formalism of Yang--Mills theory in terms of these variables is 
unfortunately somewhat involved and delicate, but for the purpose of 
the present paper, we need only note the following few points.  The 
variables $E_\mu[\xi|s]$ are nonlocal, depending on a segment of the 
parametrized loop $\xi$ around the point $\xi(s)$ on the loop labelled 
by the value $s$ of the loop parameter.  The segment has width $\epsilon$ 
which eventually is to be taken to zero, and in the limit $\epsilon \to 0$,
\begin{equation}
E_\mu [\xi|s] \to F_{\mu\nu} (\xi(s)) \dot{\xi}^\nu (s),
\end{equation}
where the dot denotes differentiation with respect to the loop
parameter $s$.  In other words, in the limit of zero segmental width,
$E_\mu[\xi|s]$ is just the Yang--Mills field at the point $\xi(s)$ dotted 
into the tangent to the loop at that point.  However, the rules of
operation are such that the limit $\epsilon \to 0$ is to be taken only
after all loop differentiations and integrations, such as that
occurring in the dual transform (\ref{dualtransf}), have already
been performed.  This generalized dual transform is thus a rather
complicated affair, but is known to reduce just to the Hodge star
for the abelian theory and in the general case to share the property 
with the Hodge star of being its own inverse apart from a sign.  A new 
feature, however, which did not occur in the abelian theory, is the matrix
$\omega (x)$ which transforms from the internal symmetry frame 
($U$-frame) in which fields of the direct formulation are measured
to the frame ($\tilde{U}$-frame) in which fields of the dual
formulation are measured.  As we shall see, this quantity will
acquire a major significance in our future discussion.  

The result of our earlier paper was that Yang--Mills theory is 
symmetric under the generalized transform (\ref{dualtransf}), and 
our present purpose is to explore the consequences.  We note first 
that this symmetry implies that in addition to the original gauge 
invariance, say $G$, the theory will possess a further gauge invariance 
(the dual invariance) $\tilde{G}$, having the same group strucutre but an 
opposite parity to the first, so that it has in all a $G \times \tilde{G}$
local gauge invariance.  Further, it implies that a dual potential
$\tilde{A}_\mu (x)$ exists which plays a role exactly dual to that
of the ordinary Yang--Mills potential $A_\mu (x)$.  Notice that
$\tilde{A}_\mu (x)$ does not represent an additional degree of
freedom to $A_\mu (x)$ since $\tilde{A}_\mu (x)$ is related to the 
dual field variable $\tilde{E}_\mu [\xi|s]$
in exactly the same way as $A_\mu (x)$ is 
related to $E_\mu [\xi|s]$, and $\tilde{E}_\mu$ is given in terms 
of $E_\mu$ via
the generalized transform (\ref{dualtransf}).  However, 
$\tilde{A}_\mu (x)$ provides an alternative description of the
gauge field to that provided by $A_\mu (x)$, and for certain phenomena,
the former may be much more convenient than the latter.  For example,
in terms of $\tilde{A}_\mu (x)$ the phase transport of the wave
function of a (colour) magnetic charge is simple, being just
$\exp i \tilde{g} \tilde{A}_\mu (x) dx^\mu$ from $x$ to a 
neighbouring point $x + dx^\mu$, whereas an expression of the same
quantity in terms of $A_\mu (x)$, though presumably possible, would
be extremely complicated.  In particular, the Wilson operator:
\begin{equation}
A(C)= {\rm Tr} \left( P \exp ig \oint_C A_\mu dx^\mu \right),
\label{ac}
\end{equation}
in the words of 't Hooft\cite{thooft}, measures magnetic flux through  
$C$ and creates electric flux along $C$.  Then by dual symmetry the 
operator:
\begin{equation}
B(C)= {\rm Tr} \left( P \exp i\tilde{g} \oint_C 
\tilde{A}_\mu dx^\mu \right)
\label{bc}
\end{equation}
should measure electric flux through $C$ and create magnetic flux
along $C$.  And indeed, using the generalized dual transform
(\ref{dualtransf}), one can show\cite{Chantsou1} that this operator 
$B(C)$ does satisfy the following commutation relation with $A(C)$, 
which was used by t' Hooft to abstractly define the $B(C)$ operator:
\begin{equation}
A(C) B(C') = B(C') A(C) \exp 2\pi in/N,  \label{comrel}
\end{equation}
where $n$ is the number of times $C'$ winds around $C$ and $N$ is
for the gauge group $SU(N)$.

Using the commutation (\ref{comrel}), 't Hooft derived the 
important result that if the electric field is confined, then the 
magnetic field is in the Higgs phase, and vice versa.  Suppose now
that $A(C)$ is confined, then $B(C)$ should be in the Higgs
phase, and its corresponding potential $\tilde{A}_\mu$ representing
the dual gauge boson should then acquire a mass and be permitted to
propagate freely through space.  At first sight, this may seem 
contradictory to the statement that $A_\mu$ (colour) is confined,
meaning that $A_\mu$ can be nonvanishing only inside hadrons,
since $A_\mu$ and $\tilde{A}_\mu$ are supposed to represent just the
same degrees of freedom.  We believe, however, that this is not the 
case.  By confinement we mean that coloured objects cannot
propagate freely in space, and a gluon $A_\mu$, being coloured, has
therefore to remain inside a hadron.  The dual gluon $\tilde{A}_\mu$,
however, is not coloured (electrically, that is).  This can be seen
in the generalized dual transform of (\ref{dualtransf}).  Under an
ordinary colour gauge (that is, in our present language, an electric
$U$-gauge) transformation $S(x)$:
\begin{equation}
E_\mu[\xi|s] \to S(\xi(s)) E_\mu[\xi|s] S^{-1} (\xi(s)).
\end{equation}
This change, however, is compensated in (\ref{dualtransf}) by a 
corresponding transformation in the matrix $\omega (x)$, which
transforms under $S(x)$ as:
\begin{equation}
\omega (x) \to \omega (x) S^{-1} (x),  \label{omega}
\end{equation}
leaving thus $\tilde{E}_\mu[\eta|t]$, and hence also $\tilde{A}_\mu$, 
invariant.  Like $\tilde{E}_\mu[\eta|t]$, $\tilde{A}_\mu$ is 
magnetically coloured but electrically colourless.   It has 
thus no reason to be confined.   And although
$A_\mu$ and $\tilde{A}_\mu$ represent the same degrees of freedom, 
specifying an $\tilde{A}_\mu$ outside hadrons in free space is not
double-counting since there $A_\mu$, by virtue of confinement, does
not propagate.  In other words, we are saying that although the gluon, 
being coloured, is confined inside hadrons, the degree of freedom it 
represents can still manifest itself in the free space outside hadrons 
as a massive, colour-magnetically charged, but colour-electrically neutral,
dual gluon.

Perhaps a more physical way of presenting the above conclusion, which
may make it easier to visualize, is to picture $\omega(x)$ itself
as a field.  It represents then a colour dyon, carrying both a colour
electric and a colour magnetic charge, transforming under 
$\tilde{U}$-transformations $\tilde{S}(x)$ as the fundamental representation
and under $U$-transformations $S(x)$ as the conjugate fundamental
representation, thus:
\begin{equation}
\omega(x) \longrightarrow \tilde{S}(x) \omega(x) S^{-1}(x).
\label{omegatransf}
\end{equation}
The dual field $\tilde{E}_\mu$ in (\ref{dualtransf})
can then be pictured as a composite object (a bound state!)
formed from an (electrically) coloured field $E_\mu$ belonging to
the adjoint representation and an $\omega$--$\bar{\omega}$ pair, in
such a way as to make the whole colour-electrically neutral,
though colour-magnetically charged.  The result is thus, in a 
sense, also a hadron, and has the right to propagate through space 
as any other hadron.  That being the case, there seems no reason
why they cannot be detected experimentally in principle.

The observation in the preceding paragraph about the matrix
$\omega (x)$ brings us to another point in duality which may
have observable consequences.  Although introduced at first by
us in all innocence as just a transformation matrix to keep
track of the gauge invariance, this $\omega (x)$ is seen to have
gradually acquired more and more physical attributes.  Thus, for
example, it was seen already in earlier papers 
\cite{Chanftsou1,Chanftsou} that in the presence
of charges, whether electric or magnetic, $\omega (x)$ will have 
to be patched.  This means that it cannot arbitrarily be put
to unity everywhere by a gauge transformation as one might expect 
for a mere transformation matrix.  Now, we find further that 
$\omega (x)$ can be combined with other fields to fundamentally
change their physical behaviour.  We propose therefore to consider
promoting $\omega (x)$ to the status of a genuine field variable.
Now in the classical field theory, $\omega (x)$ is a unitary
matrix, being an element of the gauge group.  By promoting it to a 
physical field, we mean, presumably, allowing it to fluctuate about its 
classical (vacuum) value.  We ask in such a case what physical
significance it might have.

We note first that being a transformation matrix in internal
symmetry space, $\omega (x)$ is invariant under Lorentz
transformations.  It takes a wave function for an electric
charge to one for a magnetic charge, and if we give opposite
parities to the two wave functions, as would seem natural,
the matrix $\omega (x)$ would be a space-time pseudoscalar.
Under a $U$-gauge (electric) transformation, the rows of
$\omega (x)$ transform as the conjugate fundamental representation, 
while under a $\tilde{U}$-gauge (magnetic) transformation, its 
columns transform as the fundamental representation.  Further, its 
vacuum value being a unitary matrix, its rows and columns all
have vacuum values of unit length.  In particular, then, for
an $SU(2)$ theory, a row of $\omega (x)$ would represent a
space-time pseudoscalar, isodoublet field with a vacuum
value of fixed (unit) length, as is wanted for the Higgs
field of the electroweak theory.  It is thus interesting to
entertain the possibility that the rows and columns of
$\omega (x)$ are indeed the Higgs fields in the theory 
responsible for symmetry breaking.  If this turns out to be
so, we would find ourselves in the happy position where the
Higgs fields required in the theory, which normally we have
to introduce by hand to give the desired symmetry breaking pattern,
actually arise in a natural manner as just the transformation
matrix between the direct and dual gauge frames of the theory.
It may even mean that certain aspects  in the symmetry breaking
pattern of the theory can be predicted.

The classical considerations of  our previous papers, however,
give only the vacuum configuration of the `Higgs fields' as
$\omega (x)$ but leave open the question of how exactly the
`promotion' of $\omega(x)$ to physical Higgs fields $\phi(x)$
is to be effected.  Our proposal for doing so will be given below
in section 3 when applying the idea to the standard model.  We note
that the vacuum expectation values $\omega(x)$ themselves have to 
do only with the pattern of symmetry breaking but not with the 
symmetry breaking scales.  These latter are governed by how easily 
the Higgs fields can fluctuate from their vacuum values, and by how 
rapidly these fluctuations are allowed to vary from point to point 
in space-time.  These pieces of information are encoded in standard 
formulations in the sizes of the kinetic energy term 
$\partial_\mu \phi \partial^\mu \phi$, the `mass' term $-\mu^2
\phi^2$ and in the quartic term $\lambda \phi^4$ of the Higgs action, 
relative both to one another and to the rest of the action.  These 
parameters are ultimately related to the masses of the Higgs bosons 
and the Higgsed gauge bosons, which are thus still free parameters in 
the present theory to be determined by phenomenology.  Later on, we 
shall mention some possibilities whereby duality may also help in 
constraining these parameters.

Supposing that Higgs fields can indeed be constructed in this way,
then the $\tilde{E}_\mu$ field which was pictured as a composite
formed from combining the gauge field $E_\mu$ with an
$\omega$--$\bar{\omega}$ pair can be considered as a genuine 
bound state of the gauge and Higgs fields.  Now it has already been
noted by 't Hooft \cite{thooft} that a confined system with scalar fields
in the fundamental representation of the gauge group can appear very
similar to a system in the Higgs phase, since the fundamental
`Higgses' can combine with coloured fields to form colourless bound
states which need no longer be confined.  Our picture here can thus
be regarded as just a special case of the 't Hooft scenario, in which
the naturally occurring fundamental scalar field $\omega (x)$ plays
the role of the Higgs field, and combines with the confined gluon to
give the massive, freely propagating dual gluon.  What is slightly
unusual is that both pictures here apply concurrently.

We have considered above only the pure gauge theory.  When charges are
introduced, then further consequences of duality may result.  It has 
been shown that charges in one description appear as monopoles in the 
dual description, and monopoles, being topological obstructions, 
can only have certain charges prescribed by the topology of the gauge
group.  Thus, given the electric charges of a theory, one can deduce 
what magnetic charges can occur.  Further, 't Hooft's result quoted
above implies that if the electric group is unbroken and confined,
then the dual group is broken and Higgsed, and vice versa.  Hence
given the charges we know, we have a fair idea how their dual
charges will behave.  It would therefore be interesting to enquire
whether any of these dual charges may correspond to quantum numbers
already known to us but yet unexplained.  We have in mind in
particular the question of whether the generation index which is so 
far entirely phenomenological, with no theoretical indication of its 
origin, can be interpreted as dual colour.  This last has the
advantage of occurring naturally in the gauge theory and of numbering
exactly 3, as seems indicated for the generation index by recent 
experiment.  Such questions, however, are best discussed below where 
we examine in our framework the standard model in detail.

\setcounter{equation}{0}

\section{Monopoles of the Standard Model}

We begin by collecting together some bits of information on the
standard model essential to our discussion later which though
published already in the literature \cite{Chantsou} may yet not be 
too widely known.

In most (perturbative) applications of gauge theories, one needs to
specify only the gauge Lie algebra, but for studying monopoles, one
needs also the gauge group.  Different groups may correspond to the same 
algebra.  For example, both the groups $SU(2)$ and $SO(3) = SU(2)
/{\bf Z}_2$ correspond to the same algebra $su(2)$,\footnote{We 
shall use capitals for groups but small letters for algebras.  Although
it is more correct to denote semi-simple algebras as direct sums, we
shall adhere to the product notation to avoid confusion.} and whether 
monopoles charges may exist in a theory depends on whether the gauge 
group is $SU(2)$ or $SO(3)$.

Given the gauge Lie algebra, the gauge group of a theory is to be
determined by examining what fields occur in the theory. 
\cite{Yang,Chantsou2}  For example, the maximal group generated by 
$su(2)$ is $SU(2)$, but in the pure Yang-Mills theory where only 
the gauge fields in the adjoint representation occur, 2 elements in 
$SU(2)$ differing by a sign will have the same physical effect and 
have thus to be identified.  Hence the gauge group of the theory is 
$SU(2)/{\bf Z}_2 = SO(3)$ and not $SU(2)$ itself.

An analysis along these lines taking account of all presently known 
particles and fields gives as the gauge group of the standard model 
not the maximal group $SU(3) \times SU(2) \times U(1)$ generated by 
the algebra $su(3) \times su(2) \times u(1)$, but a group obtained by
identifying the following sextets of elements in the maximal group:
\begin{equation}
(c,\!f,\!y), \!(cc_1,\!f,\!yy_1), \!(cc_2,\!f,\!yy_2), \!(c,\!ff_-,\!yy_-), 
   \!(cc_1,\!ff_-,\!yy_-y_1), \!(cc_2,\!ff_-,\!yy_-y_2),
\label{sextet}
\end{equation}
where $c, f$ and $y$ are elements respectively of $SU(3)$, $SU(2)$
and $U(1)$, with:
\begin{eqnarray}
c_r & = & \exp \frac{2 \pi i r}{\sqrt{3}} \lambda_8, r = 1, 2;
   \nonumber \\
f_- & = & \exp 2 \pi i T_3; \nonumber \\
y_r & = & \exp 4 \pi i r Y, r = 1, 2; \nonumber \\
y_- & = & \exp 6 \pi i Y.
\label{forsextet}
\end{eqnarray}
We shall call this group $U_{2,3}$, a version of $S(U(3) \times U(2))$.  
We note also that when restricted only to the electroweak sector, the 
gauge group is $U(2) = SU(2) \times U(1)/{\bf Z}_2$, and when restricted 
only to chromodynamics and electromagnetism, the gauge group is 
$U(3) = SU(3) \times U(1)/{\bf Z}_3$; in neither case is the gauge 
group the maximal group generated by the corresponding gauge Lie algebra.

The topology of the gauge group determines the values that the monopole
charges of the theory can take.  Thus, generalizing the arguments 
leading to the Dirac quantization condition for monopole charges in
electromagnetism, one can deduce in general that monopole charges are 
given by the elements of the fundamental group
$\pi_1(G)$ of the gauge group $G$.  These are the homotopy classes
of closed curves in $G$ where members of each class are curves
continuously deformable within $G$ into one another.  In particular,
for a $U(1)$ theory when the gauge group has the topology of the 
circle, $\pi_1(G) = {\bf Z}$; it follows then that monopole charges 
here are labelled by integers, namely the winding numbers around the 
circle representing $U(1)$, which is the old Dirac result.

Applied to the gauge group $U_{2,3}$, this implies that monopole 
charges of the standard model are also labelled by integers, where
a monopole labelled by $n$ can be regarded as carrying 
simultaneously:\footnote{We use here a different normalization 
convention for the gauge fields and their couplings from that used 
in our earlier publications, e.g. \cite{Chantsou}, so as to conform 
with the usual practice in the literature on the standard model.}
\begin{eqnarray}
& (a) & {\rm a \  dual \  colour \  charge} \ \zeta = \exp 2 \pi i n/3; 
   \nonumber \\
& (b) & {\rm a \  dual \  weak \  isospin \  charge} \ \eta = (-1)^n; 
   \nonumber \\
& (c) & {\rm a \  dual \  weak \  hypercharge} \ \tilde{Y} = 2 \pi n/3g_1.
\label{dualcharges}
\end{eqnarray}
Any monopole in the theory will have to carry the combination of
charges listed in (\ref{dualcharges}) for some choice of integer $n$.
We note that in (\ref{dualcharges}) dual colour and dual weak isospin
take values only in ${\bf Z}_3$ and ${\bf Z}_2$ respectively.  Thus
for dual weak isospin, $\eta = +$ corresponds to the vacuum, $\eta
= -$ to a monopole, and a monopole is its own conjugate; but for
dual colour, $\zeta = 1$ corresponds to the vacuum, while $\zeta =
\exp 2 \pi i/3$ and $\zeta = \exp 4 \pi i/3$ correspond to monopoles
of conjugate charges.

So far, one has made no use yet of dual symmetry.  For the standard
model, dual symmetry implies that in addition to the original gauge
symmetry generated by the algebra $su(3) \times su(2) \times u(1)$,
there is a further gauge symmetry generated by another algebra
$\widetilde{su}(3) \times \widetilde{su}(2) \times \tilde{u}(1)$
with the same structure but opposite parity.  Moreover, dual symmetry
says that charges in one gauge symmetry are monopoles in the dual
gauge symmetry and vice versa.  Hence, the monopole charges of 
$U_{2,3}$ listed in (\ref{dualcharges}) above can also be regarded
as ordinary (electric) charges of the dual symmetry $\widetilde{su}(3)
\times \widetilde{su}(2) \times \tilde{u}(1)$.  But charges of gauge
symmetries are usually assigned to representations of the gauge 
symmetry.  So we have to ask to what representations of the dual
symmetries the monopoles in (\ref{dualcharges}) should correspond.
The answer is as follows:
\begin{eqnarray}
(a) & \zeta = & 1 \sim {\rm \ dual \ colour \ singlet} \ {\bf \tilde{1}}, 
      \nonumber \\
    & \zeta = & \exp 2\pi i/3 \sim {\rm \ dual \ colour \ triplet} \ 
      {\bf \tilde{3}},      \nonumber \\
    & \zeta = & \exp 4\pi i/3 \sim {\rm \ dual \ colour \ antitriplet}
        \ {\bf \bar{\tilde{3}}}, \nonumber \\
(b) & \eta = & + \sim {\rm \ dual \ weak \ isospin \ singlet} \ 
      {\bf \tilde{1}},       \nonumber \\
    & \eta = & - \sim {\rm \ dual \ weak \ isospin \ doublet} \ 
      {\bf \tilde{2}},       \nonumber \\
(c) & \tilde{Y} = & n \tilde{g}_1/3.
\label{smtcharge}
\end{eqnarray}

Besides representations of the dual symmetries, the monopoles in
(\ref{dualcharges}), when considered as charges in these symmetries,
have to be further characterized by their coupling strengths, 
$\tilde{g}_3, \tilde{g}_2$ and $\tilde{g}_1$, to the dual gauge fields 
$\tilde{C}_\mu(x)$, $\tilde{W}_\mu(x)$ and $\tilde{B}_\mu(x)$ for 
respectively dual colour, dual weak isospin, and dual weak hypercharge.
Furthermore, these couplings themselves ought to be related to the 
usual colour, weak isospin, and weak hypercharge couplings $g_3, g_2$ 
and $g_1$ by conditions similar to the familiar Dirac condition 
relating the strengths of quantized electric and magnetic charges.  
The exact form of these generalized Dirac conditions depend on how 
the various quantities are normalized.  For weak hypercharge, the 
condition is the same as for electromagnetism, namely:
\begin{equation}
g_1 \tilde{g}_1 = 2 \pi,
\label{diraccond}
\end{equation}
as already implied in (\ref{dualcharges}) and (\ref{smtcharge}).  
For colour and weak isospin, if we follow the standard convention 
and write the free action as:
\begin{equation}
{\cal A}_0 = \frac{1}{4} \int d^4 x {\rm Tr} (F_{\mu\nu} F^{\mu\nu})
   + \int d^4x \bar{\psi}(i \partial_\mu \gamma^\mu - m) \psi,
\label{freeaction}
\end{equation}
with:
\begin{equation}
F_{\mu\nu} = \partial_\nu C_\mu - \partial_\mu C_\nu
   +i g_3 [C_\mu, C_\nu],
\label{Fmunu3}
\end{equation}
\begin{equation}
C_\mu = C_\mu^\alpha \lambda_\alpha/2, \ \alpha = 1, ..., 8,
\label{Cmu}
\end{equation}
for colour, and
\begin{equation}
F_{\mu\nu} = \partial_\nu W_\mu - \partial_\mu W_\nu
   +i g_2 [W_\mu, W_\nu],
\label{Fmunu2}
\end{equation}
\begin{equation}
W_\mu = W_\mu^\rho \tau_\rho/2, \ \rho = 1, 2, 3,
\label{Wmu}
\end{equation}
$\lambda_\alpha$ and $\tau_\rho$ being respectively the Gell-Mann 
and Pauli matrices, and similar formulae also for the dual quantities, 
then the generalized Dirac conditions read as follows:\cite{Chantsou1}
\begin{equation}
g \tilde{g} = 4 \pi
\label{genDirac}
\end{equation}
for both colour and weak isospin.  With these conditions the 
translation into the dual description of the information in 
(\ref{dualcharges}) on the monopole charges is now complete.

Conversely, charges in the original (direct) symmetry $su(3) \times
su(2) \times u(1)$ can also be considered as monopoles of the dual 
symmetry.  However, not knowing the experimental spectrum of the
dual charges, if any exist, we cannot as yet specify the dual gauge
group, nor yet the admissible charges its monopoles can have.  
However, if we assume that the dual gauge group is another $U_{2,3}$,
say $\tilde{U}_{2,3}$, so that its monopole charges are again given
by (\ref{dualcharges}), then it is seen that all known particles
can be accommodated as monopoles of the dual group with appropriate
choices of $\tilde{n}$.  One can thus assume without any 
inconsistency that the overall gauge group of the standard model
is $U_{2,3} \times \tilde{U}_{2,3}$, although our considerations in
what follows will not depend on this assumption.

\setcounter{equation}{0}

\section{The Promotion of $\omega$ to Higgs Fields}

We wish now to specify what we mean by promoting the rows and columns 
of the transformation matrix $\omega(x)$ to be Higgs fields.

We recall that $\omega(x)$  was originally conceived as the matrix
relating the internal symmetry $U$-frame to the dual symmetry
$\tilde{U}$-frame.  The rows of $\omega$ therefore transform as the 
conjugate fundamental representation of the $U$-symmetry, i.e. as
$\bar{\bf 3}$ of colour or $\bar{\bf 2}$ of weak isospin, while its 
columns transform as the fundamental representation of the dual 
$\tilde{U}$-symmetry, i.e. as ${\bf 3}$ of dual colour or ${\bf 2}$ of 
dual weak isospin.  Let us then introduce Higgs field $\phi^{(i)}$ and
$\tilde{\phi}^{(i)}$, with the index $(i)$ running over 1,2,3 for 
$SU(3)$ and over 1,2 for $SU(2)$, having the above transformation 
properties under respectively the $U$ and $\tilde{U}$ symmetries.

We want the vacuum expectation values of these Higgs fields to be 
such as to give an orthonormal triad for $SU(3)$ and an orthonormal
dyad for $SU(2)$, thus:
\begin{equation}
\phi^{(i)}_j(x) \longrightarrow \phi^{(i)}_0 \upsilon^{(i)}_j(x),
\label{phivac}
\end{equation}
\begin{equation}
\tilde{\phi}^{(i)}_j(x) \longrightarrow \tilde{\phi}^{(i)}_0 
   \tilde{\upsilon}^{(i)}_j(x),
\label{phitvac}
\end{equation}
where at any $x$:
\begin{equation}
\sum_k \bar{\upsilon}^{(i)}_k  \upsilon^k_{(j)} = \bar{\upsilon}^{(i)}.
   \upsilon_{(j)} = \delta^{(i)}_{(j)},
\label{upsilonn}
\end{equation}
\begin{equation}
\sum_k \bar{\tilde{\upsilon}}^{(i)}_k \tilde{\upsilon}^k_{(j)}
   = \bar{\tilde{\upsilon}}^{(i)} . \tilde{\upsilon}_{(j)} 
   = \delta^{(i)}_{(j)},
\label{upsilontn}
\end{equation}
so that we have for the transformation matrix $\omega(x)$ at any $x$:
\begin{equation}
\omega^k_j = \sum_{(i)} \upsilon^{(i)}_j \bar{\tilde{\upsilon}}^k_{(i)}.
\label{omegainups}
\end{equation}

We notice that the quantities $\upsilon^{(i)}_j$ and 
$\tilde{\upsilon}^{(i)}_j$ are actually just the frame vectors in
respectively the direct and dual description of internal symmetry
space.  In promoting them to dynamical variables as the Higgs
fields $\phi^{(i)}_j$ and $\tilde{\phi}^{(i)}_j$ as we do here is 
thus similar in spirit to the Palatini treatment of gravity in terms 
of the frame vectors or vierbeins as dynamical variables, \cite{Palatini}
with the transformation matrix $\omega$ here playing the role of the metric.

Next, if we follow the standard procedure for Higgs fields, we would 
wish presumably to obtain their vacuum expectations $\upsilon^{(i)}_j$ 
and $\tilde{\upsilon}^{(i)}_j$ as usual by minimizing some potential
${\cal V}[\phi, \tilde{\phi}]$.  Let us see what sort of a potential
we need.  First, of course, ${\cal V}$ should be invariant under the
$U$ and $\tilde{U}$-transformations given the above transformation 
properties of $\phi$ and $\tilde{\phi}$.  Second, we want ${\cal V}$
to be symmetric under permutations of the index $(i)$ of $\phi^{(i)}$
and $\tilde{\phi}^{(i)}$, given that the $\phi^{(i)}$'s for different 
$(i)$ have exactly equivalent status.  Third, given 't~Hooft's result
that if one phase is Higgsed then the dual phase is confined, we want
${\cal V}$ to be such that if $\phi_0 > 0$, then $\tilde{\phi}_0 = 0$,
and vice versa.  Fourth, given that the resulting theory should be
renormalizable, we want ${\cal V}$ to be a polynomial in $\phi$ and
$\tilde{\phi}$ of degree no higher than $4$.  Notice that although
the potential is required to be symmetric under permutations of
$\phi^{(i)}$, one expects in general that this permutation symmetry
will also be spontaneously broken along with the continuous symmetry
giving then different values to $\phi_0^{(i)}$ for different $(i)$.
Can we find an appropriate potential with such properties and yet have 
the above vacuum configuration as its minima?  

We suggest the following:
\begin{equation}
{\cal V}[\phi, \tilde{\phi}] = V[\phi] + V[\tilde{\phi}],
\label{calVex}
\end{equation}
where:
\begin{equation}
V[\phi] = -\mu \sum_{(i)} |\phi^{(i)}|^2 
   + \lambda \left\{ \sum_{(i)} |\phi^{(i)}|^2 \right\}^2
   + \kappa \sum_{(i) \neq (j)} |\bar{\phi}^{(i)}.\phi^{(j)}|^2,
\label{Vex}
\end{equation}
with the stipulation that $\mu$ is odd while $\lambda$ and $\kappa$ 
are even under the dual transform, namely $\tilde{\mu} = - \mu, 
\tilde{\lambda} = \lambda > 0, \tilde{\kappa} = \kappa > 0$.  This potential 
is interesting in that its minimum occurs (for $\mu > 0$) when the $\phi$'s
are mutually orthogonal and when $\sum_{(i)} |\phi^{(i)}|^2 = \mu/2\lambda$,
independently of the individual lengths of the different $\phi$'s.  The
minimum has thus a symmetry greater than that contained in the potential
which is only symmetric under permutations of the $\phi^{(i)}$'s, so
that different vacua from that degenerate set contained in the minimum
can be physically inequivalent.  A vacuum chosen randomly from the set 
will in general have different vacuum expectations values for all 
$\phi^{(i)}$ and we shall develop our future arguments for this general 
case for which the potential (\ref{Vex}) applies.  Our considerations 
below, however, will not depend on the explicit form of the potential.

With the above proposal as Higgs fields, let us examine the familiar 
case of electroweak symmetry breaking.  Here, of course, we know from
experiment exactly how the symmetry should be broken, namely as in 
the Salam-Weinberg manner, so that a rederivation of this result
will serve as a check on the validity of the present approach.  We
note first that in contrast to most theories, one is not allowed
here to choose whatever Higgs fields one wants to give the desired
symmetry breaking pattern but is obliged to introduce 2 (and only 2) weak 
isodoublets $\phi^{(1)}$ and $\phi^{(2)}$ as our Higgs fields.  The weak 
hypercharges of $\phi^{(i)}$, however, which have so far not entered
into our argument with $\omega$, are not yet completely specified.  
Given that our present electroweak theory has gauge group $U(2)$, as
explained in the last section, it follows that $\phi^{(1)}$ and
$\phi^{(2)}$ can only have weak hypercharges $n/2$ with $n$ being an 
odd integer (positive or negative).  However, we are still free to choose 
$\phi^{(1)}$ and $\phi^{(2)}$ having various odd half-integral values, and 
the resultant pattern of symmetry breaking will depend on the choice.

An easy way to deduce the symmetry breaking pattern for some 
given choice of hypercharges for the Higgs fields is to examine the
mass matrix for the gauge bosons arising from the kinetic energy term
in the action for the Higgs fields:
\begin{equation}
\sum_{(i)} D_\mu \phi^{(i)} D^\mu \phi^{(i)}
\label{phikinetic}
\end{equation}
where we have insisted, as in the Higgs potential, that symmetry should
be maintained between the 2 Higgs fields $\phi^{(1)}$ and $\phi^{(2)}$.
Both these Higgs field having been designated as weak isospin 
doublets, it follows that the covariant derivative is:
\begin{equation}
D_\mu = \partial_\mu -ig_2 W^\alpha_\mu(x) \frac{\tau_\alpha}{2}
   -ig_1 \frac{n}{2} B_\mu(x)
\label{covderiv}
\end{equation}
for both, each with its appropriate choice of $n$ for weak hypercharge.
The mass matrix for the gauge bosons is then given as:
\begin{equation}
W^{\alpha'}_\mu M_{\alpha' \beta'} W^{\beta' \mu}
   = \sum_{(i)} \bar{\phi}^{(i)}_V \{-g_2 W^\alpha_\mu 
   \frac{\tau_\alpha}{2} - g_1 \frac{n}{2} B_\mu\}^2
   \phi^{(i)}_V,
\label{Amassmatrix}
\end{equation}
where the primed indices $\alpha', \beta'$ on the left-hand side are 
meant to run over $1, 2, 3$ and $0$, with $W^0_\mu = B_\mu$, while the 
vacuum expection values $\phi^{(1)}_V$ and $\phi^{(2)}_V$ of the
Higgs fields on the right may be taken as:
\begin{equation}
\phi^{(1)}_V = \left( \begin{array}{c} v\\0 \end{array} \right),
   \ \phi^{(2)}_V = \left( \begin{array}{c} 0\\w \end{array} \right)
\label{phi12V}
\end{equation}
in conformity with (\ref{phivac}) and (\ref{upsilonn}) above.

From (\ref{Amassmatrix}), it is clear that $M_{\alpha' \beta'}$ is
diagonal for $\alpha'$ or $\beta' = 1, 2$, namely that 
\begin{equation}
M_{\alpha'\beta'} = \frac{g_2^2}{4} (v^2 + w^2) \delta_{\alpha'
   \beta'}
\label{A12mass}
\end{equation}
which means that the gauge bosons $W^{1,2}_\mu$ are unmixed, with 
each acquiring a mass of $(g_2/2) \sqrt{v^2 + w^2}$.  The other 2 
gauge bosons, $\alpha', \beta' = 3, 0$, on the other hand, will have
the following mass sub-matrix:
\begin{equation}
M_{\alpha' \beta'} = \frac{1}{4} \left( \begin{array}{cc}
   g^2_2 (v^2+w^2) & g_2 g_1 (n_{(1)}v^2-n_{(2)}w^2)\\
   g_2 g_1 (n_{(1)} v^2 - n_{(2)} w^2) 
   & g_1^2 (n_{(1)}^2 v^2 + n_{(2)}^2 w^2) \end{array} \right)
\label{Asubmass}
\end{equation}
for which mixing will in general occur (except when $n_{(1)} = n_{(2)}$ 
and $v = w$).  Furthermore, both the eigenstates will in general acquire 
a mass, in which case the electroweak $U(2)$ symmetry will be completely
broken.  The only situation when this will not happen is when 
$n_{(1)} = - n_{(2)} = n$, namely when the 2 Higgs field $\phi^{(i)}$
have opposite hypercharges, in which case the matrix becomes:
\begin{equation}
M_{\alpha' \beta'} = \frac{1}{4} (v^2 + w^2) \left( \begin{array}{cc}
   g_2^2 & n g_2 g_1 \\ n g_2 g_1 & n^2 g_1^2 \end{array} \right)
\label{Asubmassf}
\end{equation}
which is of rank 1, and hence has one vanishing eigenvalue.  As a 
result, the electroweak symmetry $U(2)$ will be broken down to a
residual $U(1)$ with the zero-mass eigenstate as the photon.

One sees therefore that the standard Salam-Weinberg theory does occur
as a special case of the Higgs scheme proposed above, corresponding
to the choice $n =1$ and $v = w$.  Different choices of $n$ give
different hypercharges to the Higgs fields, and the simplest choice
$n = 1$ is the one needed to give Yukawa couplings to the existing
quarks and leptons.  Different choices for $v \neq w$ on the other
hand will change only the predictions for the Higgs bosons themselves
which are not yet discovered.  Thus, as far as those of its essential
features are concerned which have so far been been tested by 
experiment, the Salam-Weinberg theory is the unique solution in the
suggested approach, so long as it is stipulated that a residual 
symmetry remains after the symmetry is broken.  It seems therefore 
that the idea of promoting the transformation matrix $\omega$ to 
Higgs fields is quite viable, having passed the test in the only 
example of spontaneous symmetry breaking in particle physics which 
has so far been confirmed by experiment.

The choice of $v = w$ would be natural if one assumes as in the 
minimal single Higgs model that $\phi^{(2)}$ is the C-conjugate of
$\phi^{(1)}$.  Here, however, we have no good theoretical reason to
make this special choice.  Although there is also nothing against
doing so, it would seem perhaps more natural in view of the proposed
interpretation of Higgs fields as frame vectors to regard $\phi^{(1)}$
and $\phi^{(2)}$ as independent fields.  In that case, there will be
more Higgs bosons, but also naturally different values for $v$ and
$w$, with the advantage of giving different masses to $u$- and 
$d$-type quarks without requiring widely different values for their 
Yukawa couplings.

\setcounter{equation}{0}
\section{The Breaking of Dual Colour}

From dual symmetry, one deduces that dual to the usual $su(3)$ clour
symmetry, there is an $\widetilde{su}(3)$ symmetry for dual colour,
and from 't Hooft's argument \cite{thooft} plus the empirical fact that 
colour is unbroken and confined, one deduces that this $\widetilde{su}(3)$
for dual colour will be broken and Higgsed.

How is the breaking of $\widetilde{su}(3)$ to be achieved?  At this 
point, we enter an uncharted domain with no longer experimental facts 
or previous experience to guide us, but according to the suggestions 
above, the symmetry breaking is to be attained by introducing as Higgs
fields 3 triplets $\tilde{\phi}^{(a)}, (a) = 1,2,3$ of dual colour
corresponding to the rows of the transformation matrix $\omega$.
These $\tilde{\phi}$'s can carry also dual or magnetic hypercharges 
which, according to the analysis in Section 2, can only take the 
values $(\tilde{n}+1/3) \tilde{g}_1$, with $\tilde{n}$ an integer 
(positive or negative), and $\tilde{g}_1 = 2 \pi/g_1$.  As with the 
breaking of the electroweak symmetry treated in the preceding section, the 
symmetry breaking pattern here will depend on what dual hypercharges 
are assigned to the Higgs fields $\tilde{\phi}^{(a)}$.

To identify the breaking pattern, let us examine again the gauge boson 
mass matrix arising from the kinetic energy of the Higgs fields:
\begin{equation}
\sum_{(a)} D_\mu \bar{\tilde{\phi}}^{(a)} D^\mu \tilde{\phi}^{(a)},
\label{phitkin}
\end{equation}
with
\begin{equation}
D_\mu = \partial_\mu - i\tilde{g}_3 \tilde{C}^\alpha_\mu 
   \frac{\lambda_\alpha}{2} - i \tilde{g}_1 (\tilde{n}_{(a)} +1/3)
   \tilde{B}_\mu,
\label{covderivdc}
\end{equation}
where $\tilde{C}_\mu$ and $\tilde{B}_\mu$ are respectively the gauge 
potentials for dual colour and dual hypercharge which, through our 
previous work\cite{Chanftsou}, we know exist.  The mass matrix for 
the gauge bosons is given as:
\begin{equation}
\tilde{C}^{\alpha'}_\mu M_{\alpha' \beta'} \tilde{C}^{\alpha' \mu}
   = \sum_{(a)} \bar{\tilde{\phi}}^{(a)}_V \left[ -\tilde{g}_3
   \tilde{C}^\alpha_\mu \frac{\lambda_\alpha}{2} - \tilde{g}_1
   (\tilde{n}+1/3) \tilde{B}_\mu \right]^2 \tilde{\phi}^{(a)}_V,
\label{massmatdc}
\end{equation}
where $\alpha', \beta' = 0, 1, ..., 8$ with $\tilde{C}^0_\mu =
\tilde{B}_\mu$, and $\tilde{\phi}^{(a)}_V$, the vacuum expectations
for the Higgs fields, can be chosen as:
\begin{equation}
\tilde{\phi}^{(1)}_V = \left( \begin{array}{c} x\\0\\0 \end{array}
   \right), \ \tilde{\phi}^{(2)}_V = \left( \begin{array}{c} 0\\y\\0
   \end{array} \right), \ \tilde{\phi}^{(3)}_V = \left( \begin{array}{c}
   0\\0\\z \end{array} \right).
\label{phitvacu}
\end{equation}
A similar analysis to that given for electroweak symmetry breaking 
in Section 3 then shows that the dual gluons $\tilde{C}^\alpha_\mu$
for $\alpha = 1,2,4,5,6,7$ remain unmixed but acquire respectively
the following masses:
\begin{eqnarray}
\alpha & = 1,2: M_\alpha & = \frac{\tilde{g}_3}{2}
   \sqrt{x^2 + y^2}, \nonumber\\
\alpha & = 4,5: M_\alpha & = \frac{\tilde{g}_3}{2}
   \sqrt{x^2 + z^2}, \nonumber\\
\alpha & = 7,6: M_\alpha & = \frac{\tilde{g}_3}{2}
   \sqrt{y^2 + z^2},
\label{massdc1}
\end{eqnarray}
whereas the other 2 components $\tilde{C}^\alpha_\mu$ for $\alpha=3,8$ 
will in general mix with each other and with the dual hypercharge 
potential $\tilde{B}_\mu$.  As for the breaking of the electroweak 
symmetry studied in Section 3, the mass matrix here  will 
leave a residual symmetry only for exceptional choices of the integers 
$\tilde{n}_{(1)}, \tilde{n}_{(2)}, \tilde{n}_{(3)}$ and of the
parameters $x, y, z$.  For example, for $\tilde{n}_{(1)} = 
\tilde{n}_{(2)} = 0, \tilde{n}_{(3)} = -1$ and $x = y = z$, there is 
a $u(1)$ symmetry left which involves both dual colour and dual 
hypercharge, but this symmetry is not the dual to electromagnetism
and has in fact no particular physical significance.  Rather, we 
want here dual colour to be completely broken, for which case there 
is a wide choice.  For lack of any other guidance, we choose to work 
with the simplest case when all $\tilde{n}_{(a)}$'s are the same 
and equal to, say, $\tilde{n}$.  In particular, we focus on 
$\tilde{n} = - 1$ giving the common dual hypercharge of all 
$\tilde{\phi}^{(a)}$ as $-2 \tilde{g}_1/3$, for, according to the 
result (\ref{dualcharges}) of our analysis in Section 2, this is 
the smallest dual hypercharge a monopole of $U_{2,3}$ can have which 
is at the same time a dual colour triplet and a dual weak isospin 
singlet as the Higgs fields $\tilde{\phi}^{(a)}$ are supposed to be.  
In that case, the mass matrix for the gauge bosons
$\tilde{C}^\alpha_\mu, \alpha = 3, 8, 0$ reads as:
\begin{equation}
\left( \begin{array}{ccc} \frac{\tilde{g}_3^2}{4} (x^2+y^2) &
   \frac{\tilde{g}_3^2}{4\sqrt{3}} (x^2 - y^2) &
   - \frac{\tilde{g}_3 \tilde{g}_1}{3} (x^2 - y^2) \\
   \frac{\tilde{g}_3^2}{4 \sqrt{3}} (x^2 - y^2) &
   \frac{\tilde{g}_3^2}{12} (x^2 + y^2 + 4 z^2) &
   - \frac{\tilde{g}_3 \tilde{g}_1}{3 \sqrt{3}} (x^2+y^2-2 z^2) \\
   - \frac{\tilde{g}_3 \tilde{g}_1}{3} (x^2 - y^2) &
   - \frac{\tilde{g}_3 \tilde{g}_1}{3 \sqrt{3}} (x^2+y^2-2 z^2) &
   \frac{4 \tilde{g}_1^2}{9} (x^2 + y^2 + z^2) \end{array} \right).
\label{massdc2}
\end{equation}
This mass matrix is a little messy to diagonalize and it is not
particularly illuminating algebraically.  We shall thus do so only 
when dealing later with numerical results.  Here, we need only note 
that since the $\widetilde{su}(3) \times \tilde{u}(1)$ symmetry is 
here completely broken, all the associated 9 gauge bosons will 
acquire nonzero masses.  

We have argued already in the Introduction that these dual colour 
and dual hypercharge gauge bosons can exist as freely propagating particles.
Their masses are unknown so long as the vacuum expectation values
$x, y, z$ of the Higgs fields $\tilde{\phi}^{(a)}$ remain undetermined
parameters.  Presumably, however, the masses will be high, at least 
in the TeV range, for otherwise the bosons would have already been
found.  Apart from possibly being observed directly as particles
in future, they can also be exchanged between dual colour and dual
hyper-charges, if such exist - and we shall be considering this possibility
in the next section - giving rise to interactions between them.  The
fact, however, that these gauge bosons are supposed to represent
just the same degrees of freedom as the colour gluons $C_\mu$ and
the hypercharge potential $B_\mu$ makes the physical effects of 
their exchange a little hard to envisage.  We suggest the following 
picture.

Consider first a pure dual colour charge of strength $\tilde{g}_3$.
It can interact with a similar charge via the exchange of a dual
gluon $\tilde{C}_\mu$.  In general, we recall, the colour 0 and 8 
components of $\tilde{C}_\mu$ will mix with one another and with the dual 
hypercharge potential $\tilde{B}_\mu$ and will not remain thus a physical
state, but for simpler presentation, let us pretend in this dicussion
that this does not happen.  The interaction looks like then that it 
will have a short range of the order of the inverse dual gluon mass 
which we have already stipulated to be large.  However, $\tilde{C}_\mu$ 
is supposed to represent just the same degree of freedom as the
colour potential $C_\mu$.  Indeed, through the dual transform 
(\ref{dualtransf}), one has in principle an explicit procedure, 
though a very complicated one, for constructing $C_\mu$ from 
$\tilde{C}_\mu$.  In particle language, this would seem to mean
that a dual gluon can transform itself, or ``metamorphose'', into
a gluon.  Will this then affect our conclusion above about the 
range of the interaction?  We think not, for the gluon, though 
massless, is confined and cannot propagate in free space, so that 
the range of the interaction will still be charaterised by the mass 
of the dual gluon.  In particular, at energies low compared with 
that mass scale, the interaction will be strongly suppressed by
the dual gluon propagator, in the same way that weak interactions
historically were considered ``weak'' in spite of its sizeable 
coupling $g_2$ because of its suppression at low energy by the
``large'' $W$ boson mass.

What will happen, however, to the interaction between dual 
hypercharges?  They will couple with strength $\tilde{g}_1$ to 
the dual hypercharge potential $\tilde{B}_\mu$ which is also 
massive, and so it looks as if the interaction will again be
short-ranged.  This, however, need not be the case for 
$\tilde{B}_\mu$, like the dual gluon above, may also metamorphose
into a $B_\mu$, but, in contrast to the colour case, $B_\mu$ has,
via electrowek mixing, a component in the photon which is not 
confined and can propagate in free space.  It seems to us therefore 
that it is eventually the photon that will govern the range of the 
interaction, but that the effective coupling is reduced from the
original $\tilde{g}_1$ to $\tilde{g}_1/M_{\tilde{g}_1}^2$ where 
$M_{\tilde{g}_1}$ is a measure of the $\tilde{B}_\mu$ mass.
We have in mind a picture as that represented symbolically in
\begin{figure}
\vspace{5cm}
\caption{Interaction of dual hypercharges}
\label{Intmagch}
\end{figure}
Figure \ref{Intmagch}, where a wavy line represents $\tilde{B}_\mu$, a 
dotted line a photon, and a little circle some sort of ``metamorphosis''
vertex.  In other words, we are suggesting that perhaps, even though
magnetic charges may exist as dual hypercharges and as such can still
interact via long-ranged Coulomb-like forces, the effective strength 
of their interaction is drastically reduced from that expected from the 
Dirac quantization value.

Later on, we shall attempt to assign dual colour and dual 
hypercharges to existing particles, in which case, for the assignment
to make physical sense, it will be essential to avoid having
unwanted forces between the particles arising out of these dual
charges.  If the picture given above for the interaction of dual
charges is correct, then one sees that the embarrassment can be
avoided by supposing sufficiently large masses for the dual colour 
and dual hypercharge gauge bosons, which is possible so long as the 
vacuum expectation values $x, y, z$ of the Higgs fields can be 
freely chosen.

\setcounter{equation}{0}
\section{Dual Colour as Generation Index}

The attractiveness to us of making dual colour into the generation
index is twofold.  On the one hand, dual colour is 3 in number, 
just like generations, and being there already in the gauge theory,
it would be surprising, as Dirac said of monopole charges when he
first discovered them, that Nature should make no use of it.
Besides, if our interpretation in the previous sections were correct,
dual colour would in any case manifest itself in a number of new
phenomena, and if it is not as generation then it has to be otherwise
accommodated.  On the other hand, from the historical point of view,
generation appears in the standard model just as an empirical concept
introduced to fit experiment.  As such it sticks out uncomfortably
in a theory which is otherwise quite geometrical, and demands from
us some understanding of its theoretical origin.

By dual symmetry, a gauge theory can be described equally in terms
of either the gauge potential or its dual.  In the usual description 
of the standard model in terms of the colour potential $C_\mu$, dual
colour charges appear as monopoles.  In Section 2, we have already
analysed what colour monopoles may occur.  Our first task therefore,
in attempting to interpret generations as dual colour, is to assign
each particle occuring in Nature a place in the table of permissible
colour monopoles.  In the preceding section, we have effectively
done so already for the gauge bosons and the Higgs fields.  We shall
try now to do the same for the fermion fields.  

For the moment, let us ignore weak isospin.  Each fermion then, 
whether quark or lepton, occurs in 3 generations.  If we wish to
identify generation with dual colour, then it would be natural to
assign the fermions to dual colour triplets, which according to 
Section 2, are permissible to colour monopoles.  Not both the left-
and right-handed fermions, however, can be assigned to dual colour 
triplets, for otherwise we would not be able to construct a Yukawa
coupling of the fermions with the Higgs fields $\tilde{\phi}^{(a)}$
introduced above which are themselves dual colour triplets.  Taking
then a hint from the Salam-Weinberg theory, let us make the 
left-handed fermions dual colour triplets but give no dual colour
charges at all to right-handed fermions, thus: $(\psi_L)_{a \tilde{a}},
(\psi_R)_a^{[b]}$ for quarks and $(\psi_L)_{\tilde{a}}, (\psi_R)^{[b]}$
for leptons, where the index $a$ denotes colour, $\tilde{a}$ dual
colour, both running from 1 to 3, while the index $[b]$, though 
also running from 1 to 3, is just a label for 3 types of dual colour
neutral right-handed fermion fields.  With this choice of dual
colour for fermions, we can then write the Yukawa coupling as:
\begin{equation}
\sum_{[b]} Y_{[b]} \sum_{(a)} (\bar{\psi}_L)^{a \tilde{a}} 
   \tilde{\phi}^{(a)}_{\tilde{a}} (\psi_R)^{[b]}_a
\label{Yucoupq}
\end{equation}
for quarks, and a similar one for leptons, where we have maintained
again, as in the Higgs potential and in the kinetic energy of the
Higgs fields, a symmetry between $\tilde{\phi}^{(a)}$ for different
$(a)$.

The mass matrix $m$ for fermions is to be obtained by inserting for
$\tilde{\phi}^{(a)}$ in (\ref{Yucoupq}) their vacuum expectations
given in (\ref{phitvacu}), giving:
\begin{equation}
m = \left( \begin{array}{ccc} xa & xb & xc \\ ya & yb & yc \\
   za & zb & zc \end{array} \right) = \left( \begin{array}{c} 
   x \\ y \\ z \end{array} \right) (a, b, c),
\label{massmatf}
\end{equation}
where we have written $a = Y_{[1]}, b = Y_{[2]}, c = Y_{[3]}$ for
short.  For this we obtain, for $\rho^2 = (|a|^2 + |b|^2 + |c|^2)$:
\begin{equation}
m m^{\dagger} = \rho^2 \left( \begin{array}{ccc} x^2 & xy & xz \\
   yx & y^2 & yz \\ zx & zy & z^2 \end{array} \right)
   = \rho^2 \left( \begin{array}{c} x \\ y \\ z \end{array} \right)
   (x, y, z),
\label{mmdagger}
\end{equation}
which being factorizable as shown, is a matrix of rank 1, having 
thus only one nonzero eigenvalue $\rho^2 \zeta^2$, with $\zeta^2 =
x^2 + y^2 + z^2$.  The matrix $m m^{\dagger}$ can be diagonalized by:
\begin{equation}
U = \left( \begin{array}{ccc} \alpha x & \alpha y & \alpha z \\
   \beta/x & \beta \omega^2/y & \beta \omega/z \\
   \gamma \left[ \frac{y \omega^2}{z} - \frac{z \omega}{y} \right] &
   \gamma \left[ \frac{z}{x} - \frac{x \omega^2}{z} \right] &
   \gamma \left[ \frac{x \omega}{y} - \frac{y}{x} \right]
   \end{array} \right)
\label{Ummdagger}
\end{equation}
with:
\begin{eqnarray}
\alpha^{-2} & = & x^2 + y ^2 + z^2, \nonumber \\
\beta^{-2}  & = & \frac{1}{x^2} + \frac{1}{y^2} + \frac{1}{z^2},
   \nonumber \\
\gamma^{-2} & = & \frac{1}{x^2 y^2 z^2} (x^2 y^4 + y^2 z^4 + z^2 x^4
   x^4 y^2 + y^4 z^2 + z^4 x^2 + 3 x^2 y^2 z^2), \nonumber \\
   & = & \alpha^2 \beta^{-2}
\label{albegam}
\end{eqnarray}
and $\omega = \exp(2 \pi i/3)$ a cube root of unity.  Thus:
\begin{equation}
U m m^{\dagger} U^{\dagger} = diag(\rho^2 \zeta^2, 0, 0).
\label{diagmmd}
\end{equation}

We conclude therefore that all the mass in this mass matrix is 
soaked up by one single massive state, leaving the other 2 massless.  
Furthermore, since the diagonalizing matrix $U$ in (\ref{diagmmd})
depends only on the vacuum expectation values $x, y, z$ of the
Higgs fields $\tilde{\phi}^{(a)}$ which are themselves independent
of which fermions they are coupled to, it follows that $U$ must be
the same for $u$-type and $d$-type quarks, giving thus the identity matrix as 
the Cabbibo-Kobayashi-Moskawa (CKM) matrix.  Now, although such a mass matrix 
is highly degenerate, it is not at all bad as a first approximation to 
the physical situation, given that for both $u$-type and $d$-type quarks 
and also for leptons, the empirical masses for the 2 lower generations are in
every case no more than 6 percent of the highest generation mass, while
the empirical CKM matrix has its diagonal elements all differing
from unity by at most 3 percent and its largest off-diagonal element of
order only 20 percent. \cite{databook}  Indeed, these significant empirical 
facts are a bit of a mystery in conventional formulations of the standard 
model, having there no obvious explanation, and we regard it as an attractive 
feature of our scheme that it should lead immediately to such a sensible
zeroth order approximation.  In the next section, we shall consider
the means whereby the above degeneracy at zeroth order may be 
lifted perturbatively to give nonvanishing values for the lower
generation masses and for the off-diagonal CKM matrix elements.
Here we only note that the masses of the highest generation, namely
$t, b$ and $\tau$, can of course  be fitted to the experimental values
by adjusting the Higgs fields vacuum expectation values $x, y, z$
and the Yukawa couplings $a, b, c$.

Obviously, the great danger in interpreting generations as a broken
gauge symmetry is that gauge symmetries imply gauge interactions, 
and none has been observed between generations besides the usual 
colour and electroweak (and of course gravitational) forces.  This is 
particularly worrying with dual colour, for the gauge interactions
here are in principle strong.  Thus, for example, the neutrinos,
which carry a generation index, and hence in the present scheme also
dual colour, can in principle interact strongly with one another,
which would be far from the truth as we now know it.  However, as
already pointed out in the last section, dual colour is broken, with
all gauge bosons acquiring masses.  The effect of their exchange 
is therefore suppressed by their propagators at energies low 
compared with their masses.  Thus, by choosing the gauge 
boson masses sufficiently high, one can in principle always
reduce the gauge interaction due to dual colour sufficiently to
keep within experimental bounds.  For example, a crude estimate
shows that by choosing dual gluon masses greater than 1 TeV, we can 
make the dual colour interactions between neutrinos in the present 
scheme weaker than the standard weak interactions between them.  Now, 
from (\ref{massdc1}) and (\ref{massdc2}), one sees that one can make
dual gluon masses as large as one likes so long as the vacuum
expectation values $x, y, z$  of the Higgs fields $\tilde{\phi}^{(a)}$
remain unconstrained. At the same time, one sees from (\ref{diagmmd})
that one can still keep the quark masses at around the experimental 
scale by adjusting appropriately the Yukawa couplings $a, b, c$.
This is the tactic we shall advocate, which is at least possible
when we are treating the breaking of dual colour in isolation from
the breaking of weak isospin as we have been doing so far in this
section.

However, in combining the treatment of symmetry breaking for both
dual colour and weak isospin, we meet with a problem.  The 
left-handed fermion is not only a triplet of dual colour but also
a doublet of weak isospin, thus: $(\psi_L)_r^{a \tilde{a}}$ for
quarks and $(\psi_L)_r^{\tilde{a}}$ for leptons while the 
right-handed fermion is a singlet in both.  Thus, given that our 
Higgs fields $\tilde{\phi}^{(a)}$ and $\phi^{(r)}$ carry each only
dual colour or weak isospin, we would need both to build an
invariant coupling with the fermion fields, e.g. for quarks:
\begin{equation}
\sum_{[b]} Y_{[b]} \sum_{(a)} (\bar{\psi}_L)_r^{a \tilde{a}}
   \tilde{\phi}^{(a)}_{\tilde{a}} \phi^{(1)r} (\psi_R)_a^{[b][1]}
   + \sum_{[b]} Y'_{[b]} \sum_{(a)} (\bar{\psi}_L)_r^{a \tilde{a}}
   \tilde{\phi}_{\tilde{a}}^{(a)} \phi^{(2)r} (\psi_R)_a^{[b][2]},
\label{mockyukawa}
\end{equation}
where the indices [1] and [2] denote the 2 types of right-handed
isosinglets with hypercharge respectively 2/3 and $-1/3$.  This is not
properly a Yukawa coupling and looks like being nonrenormalizable.

If we expand the Higgs fields $\tilde{\phi}$ and $\phi$ about 
their vacuum values, we would obtain the mass matrices of the $u$-
and $d$-type quarks respectively as:
\begin{equation}
m_u = v \left( \begin{array}{c} x \\ y \\ z \end{array} \right)
   (a,b,c),
\label{msubu}
\end{equation}
\begin{equation}
m_d = w \left( \begin{array}{c} x \\ y \\ z \end{array} \right)
   (a',b',c'),
\label{msubd}
\end{equation}
and Yukawa-type coupling to the oscillations $\tilde{\phi}'$
and $\phi'$ of the Higgs fields about their respective vacuum
expectations, thus:
\begin{equation}
\sum_{[b]} Y_{[b]} \sum_{(a)} (\bar{\psi}_L)_r^{a \tilde{a}}
   (\tilde{\phi}')_{\tilde{a}}^{(a)} v (\psi_R)_a^{[b][1]}
   + (\bar{\psi}_L)_r^{a \tilde{a}} v^{-1} m_{\tilde{a} [b]}
   (\phi')^{(1)r} (\psi_R)_a^{[b][1]}
\label{mocyukawa}
\end{equation}
for the $u$-type quarks, and a similar expression for the
$d$-type quarks.  The mass matrices (\ref{msubu}) and (\ref{msubd})
are of the form we wanted in (\ref{massmatf}) apart from a different
normalization convention.  The Yukawa couplings to the fields
$\tilde{\phi}'$ and $\phi'$ are also as expected, with the second
term being the familiar coupling to the Higgs field of standard
electroweak theory, and the first term being a coupling for the
dual colour Higgs that we can accept.  There will be, however, a
further term in the expansion in which both $\tilde{\phi}'$ and
$\phi'$ occur which, though arguably small for oscillations small
compared with their vacuum expectation values, can nevertheless
make the theory nonrenormalizable.

We have considered 2 ways of addressing this problem.  The first 
is to combine the 2 sets of Higgs fields $\tilde{\phi}^{(a)}_{\tilde{a}}$
and $\phi^{(r)}_r$ into a single set, say, $\Phi_{\tilde{a} r}^{(a)(r)}$, 
carrying both dual colour and weak isospin.  In that case, we can write 
down a genuine Yukawa coupling for our fermions as follows:
\begin{equation}
\sum_{(a)[b](r)[s]} Y_{[b][s]} (\bar{\psi}_L)^{a\tilde{a}r}
   \Phi_{\tilde{a}r}^{(a)(r)} (\psi_R)_a^{[b][s]}.
\label{genyukawa}
\end{equation}
The disadvantage, however, is that the breaking of dual colour and 
weak isospin will then be governed by the same vauum expectation
values of these Higgs fields, and hence would occur at comparable 
energy scales.  It would thus remove the freedom of pushing the
dual colour gauge bosons to high masses so as to suppress unwanted
interactions between generations as we had advocated above.  If
one takes this route, therefore, one will have to find some other
cleverer way for suppressing the unwanted interactions to
within experimental bounds, which though perhaps possible seems
to us somewhat contrived and difficult to achieve.

The alternative that we prefer which leaves free the symmetry 
breaking scale for dual colour compared with that for weak isospin
is to accept (\ref{mockyukawa}) but to regard the present scheme
as just a low energy effective theory and some of the fields we 
have so far listed as composites of some yet undiscovered more
fundamental fields.  Within the present dual framework, there is
good reason to suspect that that may indeed be the case.  Dual
symmetry implies that the electroweak $su(2)$ symmetry should have
a dual, i.e. an $\widetilde{su}(2)$ symmetry.  At the fundamental
level, therefore, one expects that Higgs fields (frames) and fermion
fields should carry also $\widetilde{su}(2)$ indices.  Up to now,
however, we have considerd only $\widetilde{su}(2)$ singlets which
are all that is required so far to accommodate the known particle 
spectrum.  The rationale for that, we suggest, is that the 
electroweak $su(2)$ symmetry being broken and Higgsed, 't Hooft's 
argument\cite{thooft} would imply that its dual $\widetilde{su}(2)$ 
should be unbroken and confined.  In that case, only $\widetilde{su}(2)$
singlets can exist in the free state, which are all that one has
seen at present, and unless one can perform deep inelastic 
experiment at high enough energy, one would not be able to see
their $\widetilde{su}(2)$ internal structure.  One can even argue
that, the $\widetilde{su}(2)$ coupling $\tilde{g}_2$, as estimated
from the experimental value of $\alpha_2 = g_2^2/4\pi \sim 0.033$
and the Dirac quantization condition (\ref{genDirac}), being more
than 10 times larger than the $su(3)$ colour coupling $g_3$, the
confinement by dual weak isospin would be much deeper than by colour 
and would require much higher energy to detect.  Now, if some of
the ``fundamental'' particles we know are in fact composites, it 
would not be surprising if some of their couplings, in particular
the ``Yukawa'' coupling (\ref{mockyukawa}), appear nonrenormalizable.
It is not easy, of course, to guess the fundamental fields and 
couplings at the deeper level, but is is not hard to find examples
which can give rise to the effective coupling (\ref{mockyukawa})
\begin{figure}
\vspace{7cm}
\caption{Example of an effective coupling}
\label{mockyukawad}
\end{figure}
we want.  The construction in Figure \ref{mockyukawad} is a
possibility, in which each line is labelled by the indices it carries,
$a$ being colour, $\tilde{a}$ dual colour, $r$ weak isospin,
$\tilde{r}$ dual weak isospin, and the last number dual weak
hypercharge.  Each line in Figure \ref{mockyukawad} is an admissible
combination of dual charges as listed in (\ref{dualcharges}) as it
ought to be.  The first (from left) and last fermion lines are the 
$\psi_L$ and $\psi_R$ above, the first Higgs line is $\phi^r$,
while the second and third Higgs lines are supposed to be confined
together by their dual weak isospin as indicated by $\tilde{r}$
to form the other Higgs fields $\tilde{\phi}^{\tilde{a}}$ as 
composites.  If the remaining fermion lines are assumed to be heavy,
we would obtain (\ref{mockyukawa}) as an effective coupling.

For the present, we leave the choice of the 2 alternatives open
as it will make no difference yet to our phenomenology, although
in the considerations which follow, our thinking may have been
biassed towards the second choice.

\setcounter{equation}{0}

\section{CKM Matrix and Masses for Lower Generations}

Although zero masses for lower generations and the identity matrix as the
CKM matrix are reasonable as zeroth order approximations, one would need of
course to envisage some mechanism whereby this degenerate scenario
can be lifted so as to give eventually more realistic values for
these parameters.  Within the framework of the standard model, loop
corrections are an obvious possibility.  However, the fermion mass
matrix here being at zeroth order factorizable as in (\ref{massmatf}),
loop corrections are quite restricted in property and it is not 
obvious at first sight that they are capable of performing that 
function.  What we wish to show now is that they can indeed do so,
at least in principle, although whether they will actually give the
correct answers to fit with experiment can only be decided by detailed
calculations.

Some one-loop corrections to the fermion mass matrix are depicted in
\begin{figure}
\vspace{7cm}
\caption{One loop corrections to the fermion mass matrix}
\label{massmatcor}
\end{figure}
Figure \ref{massmatcor}, where a full line represents fermions, a
wavy line gauge bosons, and a dotted line Higgses.  However, even
before performing any calculation, one can see that these
corrections will not alter the factorized form of the mass matrix.
Diagrams (a) and (b) will only premultiply the factorized zeroth
order mass matrix by another matrix so that the result has to remain
in the factorized form.  On the other hand, diagrams (c), (d), and 
(e) are linear combinations of matrices all of the factorized form
(\ref{massmatf}) with the same parameters $a, b$ and $c$ so that 
the result is again factorized.

A ``vertex renormalization'' diagram of the type shown in Figure
\begin{figure}
\vspace{6cm}
\caption{Vertex renormalizations to the fermion mass matrix}
\label{vertcor}
\end{figure}
\ref{vertcor}(a) can in principle break factorizability but in the 
present framework, such diagrams do not exist.  Since only the 
left-handed fermions here carry nonabelian charges ({\bf 3} for dual colour
and {\bf 2} for weak isospin) with the right-handed fermions neutral
under both these symmetries, the corresponding gauge bosons (namely
the dual gluons $\tilde{C}_\mu^{\tilde{a}}$, and the weak bosons
$W_\mu^r$) couple only to $\psi_L$, not to $\psi_R$.  $U(1)$ and 
$\tilde{U}(1)$ gauge bosons can couple to both $\psi_L$ and $\psi_R$,
depending on their $U(1)$ and $\tilde{U}(1)$ charges, but these
however do not rotate the generation (i.e. dual colour) indices,
leaving thus the factorized form of the mass matrix still intact.
On the other hand, although the diagram Figure \ref{vertcor}(b)
with a Higgs loop does exist since the Higgs couples $\psi_L$ to
$\psi_R$ as shown, the diagram has a factorized Yukawa coupling
matrix on the extreme left and right, and must therefore remain in
the factorized form.

The above analyis can be extended to diagrams with higher loops.  For
basically the same reasons as those given above for one-loop diagrams,
it can be seen that even higher loop diagrams will find it hard to 
break the factorizability of the mass matrix, and indeed we have not 
found a single one capable of doing so.  We are thus forced to accept 
that, barring non-perturbative effects, the factorized form of the 
mass matrix will remain intact to all orders.  

The fact that the mass matrix should remain factorized, however, 
does not necessarily mean that loop corrections can never lift the 
degeneracy at the zeroth order.  Take, for example, the dual gluon 
loop diagram of Figure \ref{massmatcor}(a).  Although it cannot 
break the factorizability of $m$, it will in general rotate its 
left-hand factor, thus:
\begin{equation}
m_0 = \left( \begin{array}{c} x \\ y \\ z \end{array} \right)
   (a, b, c) \longrightarrow  m_1 = \left( \begin{array}{c}
   x_1 \\ y_1 \\ z_1 \end{array} \right) (a, b, c).
\label{massmat1}
\end{equation}
The amount of this rotation will depend on the parameters in the
original zeroth order mass matrix.  In particular, these parameters 
being different for the $u$-type and $d$-type quarks, for example, 
the resultant left-hand factor $(x_1, y_1, z_1)$ after the one-loop 
correction will be different for $u$ and $d$.  It follows then 
that the matrices $U$ and $U'$ diagonalizing $m_1 m_1^{\dagger}$ 
respectively for $u$ and $d$ as given in (\ref{Ummdagger}) will
also be different, giving thus a nontrivial (i.e. non-identity) CKM 
matrix $V = U U'^{-1}$.  We notice, however, that this will happen
only when the vacuum expectation values $x, y, z$ of the Higgs 
fields are different.

As the mass matrix remains factorizable after loop corrections, it
is still of rank 1 and have thus still only 1 non-zero eigenvalue.
It might then appear that the 2 lower generations will still have
vanishing mass.  However, this need not be the case, for loop
corrections, apart from rotating the mass matrix as in 
(\ref{massmat1}), also make it run by virtue of the renormalizing
group equation, and when the mass matrix changes its value depending
on the energy scale at which it is measured, it is not immediately
clear how the actual masses of particles ought to be defined.  When
considering only one particle, the conventional wisdom is that the
running squared mass $m^2(Q^2)$ has to be evaluated at a value of 
$Q^2$ equal to its own value at that $Q^2$, which is then designated
as {\it the} mass of the particle.  When we are dealing with a mass
matrix of the factorized form (\ref{massmatf}), however, or indeed
with any matrix of rank 1, it is not so obvious what the proper 
procedure to define particle masses ought to be.  We suggest the
following.

Given that loop corrections are not supposed to break the factorized
form of the fermion mass matrix (\ref{massmatf}), it will remain of 
rank 1 at all energy scales so that the eigenstate with the highest 
eigenvalue can always be defined without any difficulty.  The other 
2 states with zero eigenvalues, however, are indistinguishable.  
Imagine then that the mass matrix is run via the renormalization 
group equation from a high energy scale down.  At every scale, we can 
diagonalize the matrix and identify the eigenstate with the nonzero 
eigenvalue.  Let us then run the scale down until this eigenvalue 
takes on the same value as the scale at which it is evaluated.  
Recalling the conventional wisdom cited above for defining the mass 
in the case of a single particle, we can then legitimately define this
value as the mass of the highest generation fermion.  At this energy,
of course, as indeed at any energy, since the mass matrix remains of
rank 1, the other 2 eigenvalues are zero, but they should not be 
interpreted as the masses of the 2 lower generations, for they are
evaluated at the wrong scale.  To find the actual masses, the mass 
matrix should be run further down in scale and evaluated at the 
masses of the lower generations, whatever these may be.  We have now
to specify exactly how this ought to be done.

The identification of the highest generation state at its mass scale
also specifies a 2-dimensional subspace of states orthorgonal to the
highest generation, namely the eigen-subspace with zero eigenvalues
in this case.  It is clear that the state vectors of the 2 lower
generations, being independent physical entities to the first, should
lie in this subspace.  Let us now run the mass matrix down to a 
lower scale.  We have seen already that loop corrections can rotate 
the left factor of the mass matrix, so that, in general, the 
$m m^{\dagger}$ matrix that we have diagonalized at the highest generation 
mass will no longer remain diagonal at the lower energy.  We can of 
course rediagonalize the matrix at the lower energy obtaining again
1 nonzero and 2 zero eigenvalues, but the diagonalizing matrix at 
the lower energy will not be the same as that at the mass of the highest
generation we have obtained before.  In other words, the 2-dimensional
subspace we have identified before at the highest generation mass scale
as containing the states of the 2 lower generations will no longer lie
within the eigen-subspace of eigenvalue $0$ at the lower energy.  To
be specific, suppose we call the eigenvector for the highest generation
${\bf v}_1$ and define 2 other mutually orthogonal (normalized) vectors
${\bf v}_2$ and ${\bf v}_3$  orthogonal also to ${\bf v}_1$, all at the
mass scale of the highest generation, then the mass submatrix:
\begin{equation}
<{\bf v}_i|m|{\bf v}_j>; i, j = 2, 3,
\label{msubmat}
\end{equation}
will in general be nonzero at the lower energy scale to which it is
run.  But this, according to the preceding arguments, has to be 
interpreted as the mass submatrix for the 2 lower generations.

The $2 \times 2$ matrix (\ref{msubmat}), being a nonzero sub-matrix of the
rank 1 matrix $m$, is of course still rank 1, so that it can also be
diagonalized at every energy giving 1 nonzero eigenvalue and the other
zero.  We can then repeat the above procedure and run the mass matrix 
on down via the renormalization group equation until the nonzero
eigenvalue of (\ref{msubmat}) equals the scale at which it is evaluated.
This value, in conformity with what has gone before, we should define 
as the mass of the second highest generation and is, of course, nonzero.

The diagonalization of the matrix (\ref{msubmat}) at the second 
generation mass identifies in turn the eigenvector with the nonzero 
eigenvalue as the state vector of the second generation.  Let us call
this vector ${\bf v}'_2$, which is by definition orthogonal to 
${\bf v}_1$, the state vector for the highest generation, and in general
different from ${\bf v}_2$.  Further, we can define the remaining
eigenvector with zero eigenvalue ${\bf v}'_3$ as the state vector of
the lowest generation, and it is by construction orthogonal to both
${\bf v}_1$ and ${\bf v}'_2$ as it should be.  At the mass scale of
the second generation, of course, the quantity ($1 \times 1$ 
submatrix):
\begin{equation}
<{\bf v}'_3|m|{\bf v}'_3>
\label{msubsubmat}
\end{equation}
vanishes, but as before, this should not be interpreted as the mass of
the lowest generation fermion since it is evaluated at the wrong scale.
We have again to run it down further via the renormalization group
equation for $m$ until the value of (\ref{msubsubmat}) equals the 
scale at which it is evaluated.  At that scale, ${\bf v}'_3$ will 
not in general lie within the eigen-subspace of $m$ with zero 
eigenvalue, so that (\ref{msubsubmat}) can be nonvanishing, or that 
the lowest generation fermion also will have nonzero mass.  Since
at each stage, the leading remaining generation soaks up all the
mass in the matrix, leaving the next generation to acquire only
whatever mass it can by running, the mass will go down by a large 
factor from each generation to the next, qualitatively the same as
what experiment is telling us.

One sees therefore that although the mass matrix remains factorizable
and of rank 1 after loop corrections, the effects of the corrections
will nevertheless be sufficient to give nonzero masses to the lower
generation fermions and to make the CKM matrix deviate from the
identity.  However, whether these effects can be made to give numbers
close to the experimental values by adjusting the free parameters
still remaining in the scheme is a question that can only be answered by a
detailed calculation, which we have begun but are far from being in
a position yet to report on.  We can at present only give the 
following two trial calculations as illustrations for the sort of
effects we shall get.

As illustration for loop corrections to the identity CKM matrix, let 
us consider for dual colour the 1-loop diagrams listed in Figure
\ref{massmatcor} which have already been evaluated by Weinberg \cite{Weinberg}
in a general Yang-Mills-Higgs framework.  He gave the answer as a
sum of 5 terms, of which the last 2 due to Higgs loops and tadpoles,
called $\Sigma_{eff}^{\phi 1}$ and $\Sigma_{eff}^{T1}$ by him, 
depend on the Higgs boson mass matrix of which we have yet insufficient 
knowledge.  The other 3 terms, depending on the Higgs fields' vacuum 
expectation values but not on their masses, we can in principle
evaluate modulo some unknown parameters and ambiguities that we shall
make clear.  Furthermore, the term called $\Sigma_{eff}^{AT}$ by
Weinberg rotates the fermion mass matrix $m$ the same way for $u$-type
and $d$-type quarks, whereas in order to contribute to the CKM matrix, a
loop correction has to rotate $m$ differently for $u$ and $d$.
There remain then only 2 terms which affect the CKM matrix directly 
for us to consider, namely:
\begin{equation}
\Sigma_{eff}^{A1} = \frac{1}{16 \pi^2} \sum_N \int_0^1 dx \left[
   -2m_W \bar{t}_N (1-x) + 4 \gamma_4 \bar{t}_N \gamma_4 m_W \right]
   \ln \left( \mu_N^2 + \frac{m_W^2 x^2}{1-x} \right) \bar{t}_N,
\label{SigmaA1}
\end{equation}
\begin{eqnarray}
\Sigma_{eff}^{A \phi} & = & \frac{1}{16 \pi^2} \sum_N \frac{1}{\mu_N^2}
   \int_0^1 dx \left\{(1-x) m_W \left[ \gamma_4 m_W, \bar{t}_N \right]
   \gamma_4 + \gamma_4 \left[ \gamma_4 m_W, \bar{t}_N \right] m_W \right\}
   \nonumber \\
   & & \left\{\ln \left(\frac{m_W^2 x^2}{1-x} \right) 
   - \ln \left( \mu_N^2 + \frac{m_W^2 x^2}{1-x} \right) \right \}
   \gamma_4 \left[ \gamma_4 m_W, \bar{t}_N \right],
\label{SigmaAphi}
\end{eqnarray}
where $\mu_N$ are the masses of the dual colour and dual hypercharge
gauge bosons, namely those listed in (\ref{massdc1}) together with the
eigenvalues of the mass matrix in (\ref{massdc2}).  The fermion mass
matrix used here is:
\begin{equation}
m_W = \frac{\rho}{\zeta} \left( \begin{array}{c} x \\ y \\ z 
   \end{array} \right) (x, y, z),
\label{massmatw}
\end{equation}
which is, crudely speaking, the square root of the matrix $m m^{\dagger}$
in (\ref{mmdagger}).  The couplings $\bar{t}_N$ are defined as:
\begin{eqnarray}
\bar{t}_N & = & -\frac{\tilde{g}_3}{2} \lambda_N \frac{1}{2} 
   (1-\gamma_5), \quad N= 1,2,4,5,6,7, \nonumber \\
\bar{t}_N & = & \left\{ -\frac{\tilde{g}_3}{2} \lambda_3 C_{3N}
   -\frac{\tilde{g}_3}{2} \lambda_8 C_{8N} + \frac{2}{3} \tilde{g}_1
   C_{0N} \right \} \frac{1}{2} (1 - \gamma_5), \ N = 3, 8, 0,
\label{tNbar}
\end{eqnarray}
with $C$ being the matrix which diagonalizes (\ref{massdc2}) and 
$\lambda_N, \ N= 1,..., 8$ the Gell-Mann matrices.

Apart from the coupling constants $\tilde{g}_3$ and $\tilde{g}_1$
which can be determined from the experimental values of their duals
$g_3 = \sqrt{4 \pi \alpha_3}$ and $g_1 = \sqrt{4 \pi \alpha_1}$
via the Dirac quantization conditions (\ref{diraccond}) and
(\ref{genDirac}), the expressions in (\ref{SigmaA1}) and 
(\ref{SigmaAphi}) depend on the vacuum expectation values of the
Higgs fields $x, y, z$ and on the Yukawa couplings $a, b, c$ 
through $\rho$, as defined in section 5.
The parameters $x, y, z$ are unknown, but once these are given then
$\rho$ can in principle be determined by normalizing $\rho \zeta$, 
the nonzero eigenvalue of $m$, on the experimental mass of the
highest generation fermion, namely $m_{top}$ and $m_{bottom}$ for
respectively the $u$-type and $d$-type quarks.  In practice,
however, there is here an ambiguity in normalizing $\rho$ for the
following reason.  There are terms in $\Sigma_{eff}^{(A1)}$ as 
well as in the other Weinberg terms that we have dropped which are
scale dependent, and though either not rotating $m_W$ at all, or
else rotating $m_W$ the same way for $u$- and $d$-type quarks, and
so not affecting the CKM matrix directly, nevertheless changes the
normalization of $m_W$.  This is presumably related ultimately to
the running of these quantities with changing scales which we have
not yet sorted out fully.  As a result, we have to treat $\rho$ also as
a parameter for the moment, and cannot fix the actual size of
off-diagonal CKM matrix elements.  Further, not having sorted out
the running effects, we also cannot, using the method outlined
earlier in this section, identify the quarks of the 2 lower 
generations.  Hence, we cannot at present specify $V_{us}$ and 
$V_{cd}$, or distinguish $V_{ub}$ from $V_{cb}$ and $V_{td}$ from 
$V_{ts}$.  The significance of this present exercise is thus 
strictly limited.

Putting in arbitrarily the parameters $x = 1, y = 2/3, z = 1/3$, we
obtained from (\ref{SigmaA1}) and (\ref{SigmaAphi}) the following
matrix for the absolute values of CKM matrix elements, where $\rho$ has 
been adjusted to give off-diagonal elements roughly of the order of 
a percent.  
\begin{equation}
\left( \begin{array}{ccc} .9998 & .0173 & .0130 \\
                          .0166 & .9998 & .0124 \\
                          .0130 & .0123 & .9998  \end{array} \right).
\label{mockCKM}
\end{equation}
Given the limitations stated in the preceding paragraph, the only 
conclusions we can draw at present are that this `mock' CKM matrix (i) does 
get rotated from the identity by loop corrections, (ii) remains roughly 
though not exactly symmetric, and (iii) is in general complex, all of
which are properties apparently exhibited by the actual CKM
matrix obtained from 
experiment. \cite{databook}  This is not much, but still enough perhaps 
as encouragement for further exploration.

As illustration for generating masses for lower generation fermions,
consider the renormalization group equations usually given for the
standard model \cite{Grzad}:
\begin{equation}
16 \pi^2 \frac{dU}{dt} = \frac{3}{2} (UU^{\dagger} - DD^{\dagger}) U
   + (\Sigma_u - A_u) U,
\label{dUdt}
\end{equation}
\begin{equation}
16 \pi^2 \frac{dD}{dt} = \frac{3}{2} (DD^{\dagger} - UU^{\dagger}) D
   + (\Sigma_d - A_d) D,
\label{dDdt}
\end{equation}
where $U$ and $D$ are respectively the Yukawa coupling matrices to
the electroweak Higgs field for respectively the right-handed $u$-
and $d$-type quarks\footnote{The symbol $U$ adopted here following 
the usual convention should not, of course, be confused with the
diagonalizing matrix in (\ref{Ummdagger}).}, and $\Sigma_{u,d}$ and 
$A_{u,d}$ are the Higgs self-energy and gauge boson loop contributions 
whose explicit forms need not here bother us.  

The matrices $U$ and $D$ can of course be diagonalized at any scale, 
but do not remain diagonal in general on running, and what interest
us for the problem at hand are just those terms which  contribute
towards the de-diagonalization of $U$ and $D$, namely the 
$DD^{\dagger}$ term in (\ref{dUdt}) and the $UU^{\dagger}$ term in
(\ref{dDdt}).  In the basis where $U$ is diagonal, $D$ is not
diagonal, and vice versa, by virtue of a nontrivial CKM matrix $V$,
so that for the de-diagonalizing effects alone which interest us,
we may write the renormalization group equations (\ref{dUdt}) and
(\ref{dDdt}) as:
\begin{equation}
16 \pi^2 \frac{dU}{dt} = -\frac{3}{2} DD^{\dagger} U,
\label{dUdta}
\end{equation}
\begin{equation}
16 \pi^2 \frac{dD}{dt} = -\frac{3}{2} UU^{\dagger} D.
\label{dDdta}
\end{equation}

Now in the philosophy of the present scheme, the main effect for
de-diagonalizing $U$ and $D$ is supposed to come from diagrams
with dual colour gauge and Higgs boson loops, as already discussed 
above.  These dual colour loop effects, however, have not been 
included in the equations (\ref{dUdt}) and (\ref{dDdt}), which 
indeed we do not even yet know how to calculate.  However, since
it was these omitted effects which are supposed to give rise to 
the nontrivial CKM matrix in the first place, the de-diagonalizing 
effects from the mixing due to the CKM matrix itself which are 
included in (\ref{dUdta}) and (\ref{dDdta}) would have to be 
regarded in this philosophy as only secondary effects induced by 
the primary dual colour loop contributions.  Nevertheless, we think 
it worthwhile to study (\ref{dUdta}) and (\ref{dDdta}) as 
illustrations for the effects on the lower generation fermion masses 
that one can expect.

As we shall be interested in running the equations only over small
ranges of the order of the mass differences between generations, we
may take the linearized equations and consider the CKM matrix itself
as constant over these ranges.  Starting then with a diagonalized
mass matrix at the mass scale of the highest generation, in our case
$diag(m_{top}, 0, 0)$ for the $u$-type and $diag(m_{bottom}, 0, 0)$
for the $d$-type quarks, and running it down to lower energies, we 
obtain:
\begin{equation}
U_t = V diag \left(\exp\left[-\frac{3(m_b/w)^2 t}{32 \pi^2} \right],
   0, 0 \right) V^{-1} diag(m_t, 0, 0),
\label{Urunning}
\end{equation}
\begin{equation}
D_t = V^{-1} diag \left(\exp\left[-\frac{3(m_t/v)^2 t}{32 \pi^2} \right],
   0, 0 \right) V diag(m_b, 0, 0),
\label{Drunning}
\end{equation}
where, $V$ being non-diagonal, one sees that the mass matrices, though
diagonalized at the highest generation mass, will become non-diagonal 
when run to the lower energy, as expected.

Now in the philosophy of the present scheme, the difference in the
top and bottom mass comes mainly from the the difference between
$v$ and $w$, i.e. the vacuum expectation values of resepectively the 
Higgs fields $\phi^{(1)}$ and $\phi^{(2)}$, so that the Yukawa
couplings $m_t/v$ and $m_b/w$ are comparable in magnitude.  In that
case, we can put:
\begin{equation}
m_t/v \sim m_b/w \sim (180 GeV/246GeV).
\label{Yucoups}
\end{equation}
Inserting this value in (\ref{Urunning}) and (\ref{Drunning}) above,
together with the experimentally measured values of the CKM matrix
elements, one obtains that on running from the highest generation to 
the next, say e.g. from the top (bottom) to the charm (strange) quark mass, 
the equations would generate off-diagonal elements in $U$ or $D$ of the 
order of $10^{-3}$ times the highest generation mass.  This is not enough to 
explain the actual mass values of the second generation which is
of the order of a few percent of the highest generation.  However,
one recalls that the effect represented by (\ref{Urunning}) and
(\ref{Drunning}) are supposed to be only secondary effects obtained
from the primary dual colour effects that we have not yet learned
to calculate.  If we argue naively that the factor of suppression 
in mass from one generation to the next due to the primary effect 
should be of the order of the square root of that due to the secondary 
effect, then the answer we obtained is about right.  The above argument,
for whatever it is worth, can be repeated for the suppression from 
the second to the lowest generation and the answer is still comparable 
with what is seen in experiment.

\section{Concluding Remarks}

The feature we find most attractive in the present scheme is the
possibility to assign both to the Higgs fields and to the fermion 
generations each a natural place.  This is a consequence of the recently
discovered nonabelian dual symmetry\cite{Chanftsou} to the extent 
that the necessary niches exist because of it in the form of the 
transformation matrix $\omega$ and of the concept of a local dual 
colour symmetry.  But the actual assignment of these niches to Higgs 
fields and to fermion generations involves of course some, perhaps 
somewhat daring but to us quite reasonable, assumptions and the 
merit or otherwise of these must rest in the end on the compatibility 
of their predictions with experiment.

As far as present investigations go, the scheme has scored a number
of positive points, among which we count the prediction of exactly 
3 generations, the mass hierarchy between them, the near identity CKM
matrix, and the possibility of evaluating lower generation masses
and off-diagonal CKM matrix elements perturbatively.  The first 
three points are all significant and noted empirical facts which
lack explanation in the usual formulation of the standard model, but
seem to have found each a {\it raison d'etre} in the present scheme.

On the other hand, there are also consequences which can give rise
to potential disagreement with experiment, among which the most
worrying is the prediction of new interactions due to dual gluon
exchange.  We argued above that these are suppressed by the dual
gluon propagator, and so long as these are large enough, we may 
not notice the interactions due to their exchange at present
experimental energy.  This suppression, however, has its limit,
on 2 counts.  First, the loop corrections, which we claimed in the
preceding section may lead to nonzero off-diagonal CKM matrix
elements and lower generation masses, also depend on the masses of
dual gluon, and if one makes these latter masses too large, then the 
loop correction may be too small to explain the experimental effects.
One shall then have to devise other means for lifting the zeroth 
order degeneracy.  Secondly, even if one can make the masses of 
dual gluons very large, there will eventually come a point at which 
the propagator suppression will no longer work, and the interaction from 
dual gluon exchange, say e.g. between neutrinos which carry dual colour,
will become very strong.  Will this not violate some astronomical
or cosmological bounds?  We do not know.  By the same token, the 
scheme may conceivably be in conflict with some currently 
held theoretical ideas on asymptotic behaviour.  At first sight, 
it may appear that the dual colour coupling $\tilde{g}_3$, being 
inversely proportional to the usual colour coupling $g_3$, will 
grow with energy and so spoil completely such cherished concepts as 
asymptotic freedom.  We are, however, not sure that this will be so.  
As already stated repeatedly above, the dual gluon $\tilde{C}_\mu$ 
does not represent a different degree of freedom to the gluon 
$C_\mu$, but should rather be regarded as a composite (a hadron!) 
formed from the usual colour gluons and the coloured Higgs fields.  
If so, their exchange should be compared not with elementary 
exchanges but with say pion-exchange between hadrons which do not 
spoil asymptotic freedom.  Nevertheless, at finite energies, dual 
gluon exchanges will affect the running of various quantities and 
hence may lead to potential discrepancy with experiment.

Let us assume optimistically that the present scheme will survive
these possible pitfalls, either as it is here proposed or with some 
modifications utilising some of the freedom still available.  We
shall find it interesting then to note that it has also some predictions 
which are probably accessible to experimental tests in the not too distant 
future.  There are first the dual gauge bosons and dual coloured Higgses.
Crude estimates from our trial calculation of the CKM matrix reported 
in Section 6 suggest that dual gauge bosons may have masses in the 
several Tev range, and if so may be accessible to LHC.  As for the masses 
of the dual coloured Higgses, however, we have at present no idea of 
their magnitudes.  Secondly, there is the exciting possibility suggested 
at the end of Section 5 that there may be yet a deeper level of confinement 
than colour with dual weak isospin. If so, future deep-inelastic experiment at 
ultra-high energy may reveal internal structures to what are presently 
regarded as elementary objects such as quarks and leptons.

Finally, we remark that dual symmetry is claimed to be inherent to
Yang-Mills theory as it is to electromagnetism.  If this is true,
then its effects would be unavoidable, and even if one does not choose
to interpret the internal symmetry frames
as Higgs fields and dual colour as generation as
we do here, the existence in theory of these niches as consequences 
of dual symmetry on the one hand, and the empirical requirement of 
Higgs fields and fermion generations on the other, would still have 
to be accounted for in some manner.

\vspace{1cm}
\noindent{\Large\bf Acknowledgement}\\

We are grateful to Bill Scott for teaching us some rudiments of the
CKM matrix and to Dick Roberts, Ben Allanach, and Herbi Dreiner for
helpful conversations on practicalities in running the fermion mass 
matrix.  One of us (TST) thanks the Wingate Foundation for partial
support.


\begin{thebibliography}{99}

\bibitem{Dirac} P A M Dirac, Proc. Roy. Soc. London {\bf A133}, 60, (1931);
   Phys. Rev. {\bf 74}, 817 (1948).

\bibitem{Schwinger} Julian Schwinger, Phys. Rev. {\bf 144}, 1087, (1966);
   Phys. Rev. {\bf D12}, 3105, (1975).

\bibitem{Zwanziger} D. Zwanziger, Phys. Rev. {\bf 176} 1480, (1968);
   Phys. Rev. {\bf D6}, 458, (1972).

\bibitem{Wuyang} Tai-Tsun Wu and Chen-Ning Yang, Phys. Rev. {\bf D12}, 3845,
   (1975); Phys. Rev. {\bf D14}, 437, (1976).

\bibitem{Monlive} C. Montonen and D. Olive, Phys. Lett. {\bf 72B}, 117,
   (1977).

\bibitem{Deser} S. Deser, J. Phys. {\bf A15}, 1053, (1982).

\bibitem{Lubkin} E. Lubkin, Ann. Phys. (New York) {\bf 23}, 233, (1963).
 
\bibitem{Coleman} S. Coleman, in {\it New Phenomena in Subnuclear Physics},
   ed. by A. Zichichi, (Plenum, New York, 1976), p. 297.

\bibitem{Halpern} M. Halpern, Nucl. Phys. {\bf B139}, 477, (1978).

\bibitem{thooft} G. 't Hooft, Acta Physica Austrica suppl. XXII, 531, (1980).

\bibitem{Chanstsou} Chan Hong-Mo, P. Scharbach, and Tsou Sheung Tsun,
   Ann. Phys. (New York) {\bf 167}, 454, (1986).

\bibitem{Seiten} N. Seiberg and E. Witten, Nucl. Phys. {\bf B426}, 19
   (1994); {\bf B431}, 484 (1994).

\bibitem{Seiberg} K. Intriligator and N. Seiberg, Nucl. Phys. {\bf B431}, 551
   (1994); N. Seiberg, {\it ibid.} {\bf B435}, 129 (1995); N. Seiberg, in
   {\it Particles, Strings, and Cosmology}, Proceedings of the Workshop,
   Syracuse, New York, 1994, edited by K.C. Wali (World Scientific, Singapore,
   1995).

\bibitem{Witten} C. Vafa and E. Witten, Nucl. Phys. {\bf B431}, 3 (1994);
   E. Witten, Math. Res. Lett. {\bf 1}, 769 (1994).

\bibitem{Aharony} O. Aharony, Phys. Lett. {\bf 351B}, 220 (1995); 
   L. Girardello, A. Giveon, M. Porrati, and A. Zaffaroni, Nucl. Phys. 
   {\bf B448}, 127 (1995);
   A. Ceresole, R. D'Auria, S. Ferrara and A. Van Proeyen,
   Nucl. Phys. {\bf B444} 92 (1995);
   J.A. Harvey, G. Moore and A. Strominger, Phys. Rev {\bf D52} 7161,
   (1995); 
   R.G. Leigh and M.J. Strassler, Nucl. Phys. {\bf B447}, 95 (1995); 
   E. Comay Nuovo Cim. {\bf B110}, 1347, (1996); 
   D. Kutasov and A. Schwimmer, Phys. Lett. {\bf 354B}, 315 (1995); 
   K. Intriligator and P. Pouliot, Phys. Lett. {\bf 353B}, 471 (1995);
   K. Intriligator, Nucl. Phys. {\bf B448}, 187 (1995); 
   C. Gomez and E. Lopez, Phys. Lett. {\bf 357B}, 558 (1995);
   E. Witten, hep-th/9505186; 
   G.W. Gibbons and D.A. Rasheed, Nucl. Phys. {\bf B454}, 185 (1995); 
   K. Intriligator, R.G. Leigh and M.J. Strassler, Nucl. Phys. {\bf B456},
   567 (1995); 
   E. Alvarez, L. Alvarez-Gaum\'e and I. Bakas, Nucl. Phys. {\bf B457}, 3 
   (1995); 
   Y. Lozano, Phys. Lett. {\bf B364}, 19 (1995);
   C. Ford and I. Sachs, Phys. Lett. {\bf B362}, 88 (1995);
   O. Aharony and S. Yankielowicz, Nucl. Phys. {\bf B473}, 93, (1996);
   Kimyeong Lee, E.J. Weinberg and Piljin Yi, Phys. Lett. {\bf B376}, 
   97, (1996);
   S.F. Hewson and M.J. Perry, hep-th/9603015.   
   P.J. Hodges and N. Mohammedi, hep-th/9608023;
   F. Ferrari, hep-th/9611012;
   J.H. Brodie and M.J. Strassler, hep-th/9611197.
   The above is still only a partial list of the many articles which 
   have recently appeared on the subject.

\bibitem{Chanftsou1} Chan Hong-Mo,
   J. Faridani and Tsou Sheung Tsun, Phys. Rev. {\bf D51}, 7040, (1995);
   Phys. Rev. {\bf D53}, 7293, (1996).

\bibitem{Chanftsou} Chan Hong-Mo, J. Faridani, and Tsou Sheung Tsun,
   Phys. Rev. {\bf D12}, 7293, (1996).

\bibitem{Chantsou1} Chan Hong-Mo and Tsou Sheung Tsun, in preparation.

\bibitem{Chantsou} For a review, see e.g. Chan Hong-Mo and Tsou Sheung 
   Tsun, {\it Some Elementary Gauge Theory Concepts} (World Scientific, 
   Singapore, 1993).

\bibitem{Yang} Chen Ning Yang, Phys. Rev. {\bf D1}, 2360, (1979).

\bibitem{Chantsou2} Chan Hong-Mo and Tsou Sheung Tsun, Nucl. Phys.
   {\bf B189}, 364, (1981).

\bibitem{Palatini} See e.g. F.W. Heyl et al., Rev. Mod. Phys. {\bf 48},
   393, (1976).

\bibitem{databook} See e.g. {\it Review of Particle Physics} ed.
   R.M. Barnett et al. Phys. Rev. {\bf D54}, 1, (1996) as summarized
   in the data booklet.

\bibitem{Weinberg} Steven Weinberg, Phys. Rev. {\bf D7}, 2887, (1973).

\bibitem{Grzad} See e.g. B. Grzadkowski, M. Lindner and S. Theisen,
   Phys. Lett. {\bf B198}, 64, (1987).


\end{thebibliography}
\end{document}